# UWarp: A Whole Slide Image Registration Pipeline to Characterize Scanner-Induced Local Domain Shift

# Authors

Antoine Schieb [1], Bilal Hadjadji [1], Natalia Fernanda Valderrama [1], Daniel Tshokola Mweze [1], Valentin Derangère [2], Laurent Arnould [2], Sylvain Ladoire [2], Alain Lalande [3,4], Alessio Fiorin [5,6,7], Carlos López Pablo [5,6,7], Noèlia Gallardo Borràs [5,6,8], Shrief Abdelazeez [9,10], Vincenzo Della Mea [11], Anna Korzynska [9], Nathan Vinçon [1], Louis-Oscar Morel [1]

[1] Ummon HealthTech SAS, 21000 Dijon, France

[2] Department of Medical Oncology, Centre Georges-François Leclerc, 21000 Dijon, France

[3] ICMUB laboratory, Université de Bourgogne Europe, 21000 Dijon, France

[4] Department of Medical Imaging, University Hospital of Dijon, 21000 Dijon, France

[5] Department of Pathology, Hospital de Tortosa Verge de la Cinta (HTVC), Catalan Health Institute (ICS), C/ Esplanetes no. 14, 43500 Tortosa, Spain

[6] Oncological Pathology and Bioinformatics Research Group, Pere Virgili Institute for Health Research (IISPV), C/ Esplanetes no. 14, 43500 Tortosa, Spain

[7] Department of Computer Engineering and Mathematics, Universitat Rovira i Virgili (URV), Tarragona, Spain

[8] Department of Basic Medical Sciences, Universitat Rovira i Virgili (URV), Tarragona, Spain

[9] Laboratory of Analysis and Processing of Microscopic Image, Nałęcz Institute of Biocybernetics and Biomedical Engineering Polish Academy of Sciences, Ks. Trojdena 4, 02-109 Warsaw, Poland

[10] System and Biomedical Engineering Department, Faculty of Engineering, Cairo University, Giza, Cairo, Egypt

[11] Department of Mathematics, Computer Science and Physics, University of Udine, Udine, Italy

# Abstract

Histopathology slide digitization introduces scanner-induced domain shift that can significantly impact computational pathology models based on deep learning methods. In the state-of-the-art, this shift is often characterized at a broad scale (slide-level or dataset-level) but not at the patch-level, which limits our comprehension of the impact of localized tissue characteristics on the accuracy of the deep learning models. To address this challenge, we present a domain shift analysis framework based on UWarp, a novel registration tool designed to accurately align histological slides scanned under varying conditions. UWarp employs a hierarchical registration approach, combining global affine transformations with fine-grained local corrections to achieve robust tissue patch alignment. We evaluate UWarp using two private datasets, CypathLung and a dataset from the BosomShield project, containing whole slide images scanned by multiple devices. Our experiments demonstrate that UWarp outperforms existing open-source registration methods, achieving a median target registration error (TRE) of less than 4 pixels (<1 µm at 40× magnification) against 21 pixels (~ 5 µm at 40x magnification) for the best performing state-of-the-art algorithm while significantly reducing computational time. Additionally, we conduct a localized characterization of scanner-induced domain shift by analyzing the predictions of a deep learning model for breast cancer pathological response prediction. Accurate patch-level alignment between scanners is performed using UWarp. Our analysis reveals that prediction variability is correlated to tissue versus background ratio and nuclei density at the patch level, highlighting the importance of accounting for localized domain shifts to improve model robustness and domain adaptation strategies in computational pathology.

# Highlights

- UWarp enables accurate WSI registration across different scanner conditions
- Combines global and local corrections for hierarchical slide alignment
- Achieves <1 µm TRE at 40× magnification, outperforming state-of-the-art methods
- Enables patch-level domain shift analysis in histopathology slides
- Prediction variability linked to tissue proportion and nuclei density in patches

# Keywords

# 1.Introduction

Artificial intelligence is transforming healthcare by enabling faster, cheaper and more accurate solutions for diagnosing, treating, and preventing diseases. In computational histopathology, deep learning has become a powerful tool for automating tissue image analysis, improving diagnostic accuracy, and detecting patterns, morphological features, and biomarkers that support disease diagnosis and prognosis [28].

Domain shift refers to the differences in data distributions between training and inference environments which can lead to a significant decline in AI model performance. In digital pathology, scanner-induced variations—such as differences in staining protocols, scanning resolutions, and imaging hardware—can cause domain shift. These variations affect model generalization, resulting in misclassifications and reduced diagnostic accuracy. Since AI models are often trained on data from specific institutions or scanners, their performance can deteriorate when exposed to unseen domain variations, limiting their robustness in clinical applications. Addressing and understanding this variability is a key challenge in developing safe and reliable AI tools for histopathology, as it directly impacts their robustness, clinical validity, and trustworthiness in real-world applications.

A popular approach to overcoming domain shift in histopathology consists in data augmentation, which can be achieved through regular image transformations (reflections, brightness and color adjustments, etc. ), stain augmentation [1], or generative neural networks [2] to produce synthetic images for training. Other approaches like stain-normalization [3–5], style transfer models [6], or feature space alignment aim to generate domain-agnostic feature representations. However, despite these efforts, model generalization often remains limited in real-world scenarios. This inconsistency raises a critical question: why do state-of-the-art techniques aimed at mitigating variability still fail to guarantee robust performance? A key reason may lie in the lack of precise, localized evaluation of domain shift. Most existing methods operate at the dataset or slide level, relying on synthetic transformations or statistical proxies that do not fully reflect the fine-grained variability present in actual tissue images.

To address this challenge, recent studies have examined domain shift at broader scales, including datasets and whole slides, by analyzing activation differences in convolutional neural networks (CNNs) to compare feature distributions between source and target datasets [7–9]. While these methods offer insight into overall domain variability, they lack the granularity needed to identify localized differences at the patch level. This limitation could potentially restrict our ability to pinpoint factors affecting model performance.

Localized (patch-level) domain shift and its effect on latent space representations has been indirectly explored through CycleGAN-based stain translation methods [7]. However, these techniques depend on synthetic image generation, which may not accurately represent real tissue image variability. As noted by Nisar et al.[7], CycleGAN models can introduce subtle, imperceptible noise that alters the data distribution, potentially leading to misleading conclusions about domain shift. This underscores the need to directly investigate patch-level domain shift using real images rather than synthetic approximations.

A promising approach for examining natural domain shift at the patch level is to retrieve and compare the same tissue region across different scans. This process involves registering same-stain scans of the same tissue slice to establish a geometric transformation that aligns the tissue area between the source and target coordinate systems. A similar idea has been explored by the authors of The PLISM dataset [10], a multi-stain and multi-device histopathology dataset which contains registered WSIs, and aligned patches with manual validation of the registrations. The authors focused on assessing the texture and color differences between the different domains instead of directly running the predictions of a task-specific model on the registered patches.

Thus, this paper introduces an approach for understanding and characterizing scanner-induced domain shift in digital pathology. By precisely aligning whole slide images at the patch level across different scanners, registration enables direct comparisons of corresponding tissue regions, facilitating a localized analysis of domain variability. Our method allows researchers to study natural domain shift in real tissue samples rather than relying on synthetic approximations, providing deeper insights into the factors influencing AI model performance.

Existing registration methods for histopathology often focus on reconstructing tissue volumes from serial sections or aligning multi-stain images for spatial analysis. These approaches vary from rigid affine transformations to non-rigid methods based on deformation fields, and from classical feature-based algorithms like SIFT to recent deep learning models. However, no methods are tailored for high-precision registration of same-stain, same-slice whole slide images scanned under different

conditions—despite this being a critical need for studying scanner-induced domain shift. While these methods perform reasonably well on same-stain slides, they can be computationally expensive, often unstable against non-rigid artifacts, and rarely achieve cell-level accuracy. Notably, Theelke et al. [13] proposed an iterative multi-resolution framework while VALIS [22] offers high-end alignment for multiplexed imaging but comes with substantial computational requirements. In contrast, our method specifically targets intra-slide, inter-scanner variability with a design optimized for patch-level alignment accuracy and efficiency.

To overcome these limitations, we present UWarp, a novel framework for registering same-stain, same-slice histopathology images with high precision and computational efficiency. Our framework introduces a precise iterative patch-based registration method that automatically generates and matches landmarks across tissue samples. This process begins at low resolution and incrementally refines alignment to the highest resolution. Matched landmarks are rigorously filtered using a quality score, and a global linear transformation function is computed using the least squares algorithm, with local non-rigid adjustments applied to account for potential misalignments or distortions introduced during the scanning process.

To summarize, this paper makes two key contributions. First, we propose a novel approach to understanding and characterizing localized scanner-induced domain shift in digital pathology by leveraging slide registration. This method facilitates direct comparisons of corresponding tissue regions across different scans, providing valuable insights into localized domain variability setting up a high quality standard. Second, to enable accurate implementation of the first contribution, we introduce UWarp, a high-precision registration technique specifically designed for same-stain, same-slice histopathology images. UWarp outperforms state-of-the-art methods in terms of accuracy and computational efficiency, establishing it as a robust solution for histopathology image analysis.

# 2.Methods

The proposed method for analyzing and characterizing localized scanner-induced domain shifts in digital pathology is illustrated in **Figure 1**. This approach relies on a registration pipeline to automatically align corresponding tissue patches between source and target scanners. By ensuring precise alignment, the same predictive deep learning model can be consistently applied to tissue regions from different scans, enabling a detailed assessment of local predictive variations due to inter-scanner variability.

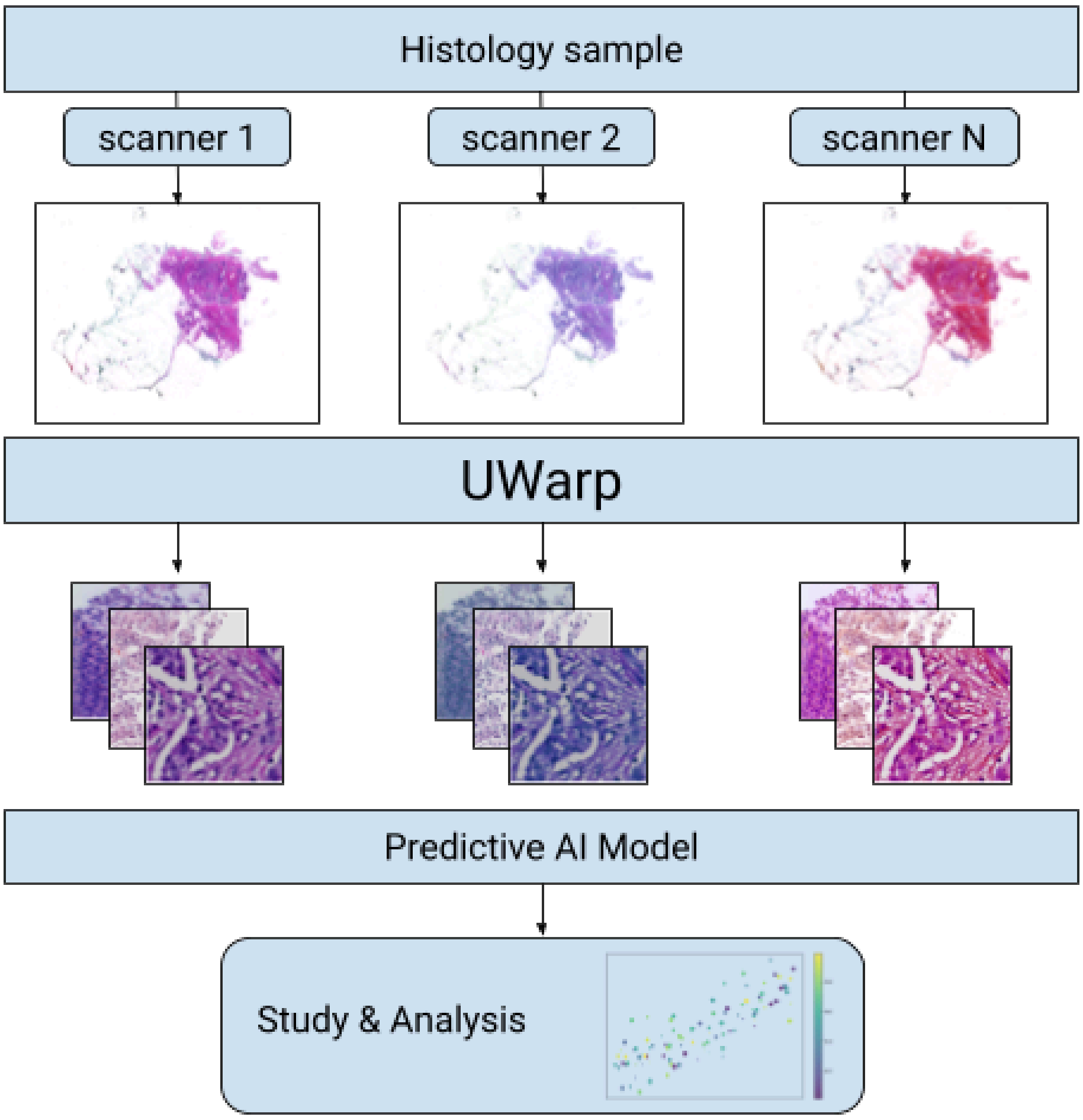

*Figure 1: Overview of the domain shift study protocol using UWarp.*

We introduce UWarp, a registration method designed to accurately align tissue patches across whole slide images acquired from different scanners. As illustrated in **Figure 2**, the pipeline comprises the following steps: (1) initial alignment, (2) automatic selection of source landmarks, (3) iterative patch-level registration to identify corresponding target landmarks, (4) computation of a global affine transformation, and (5) application of local non-rigid adjustments. Each component is described in detail below.

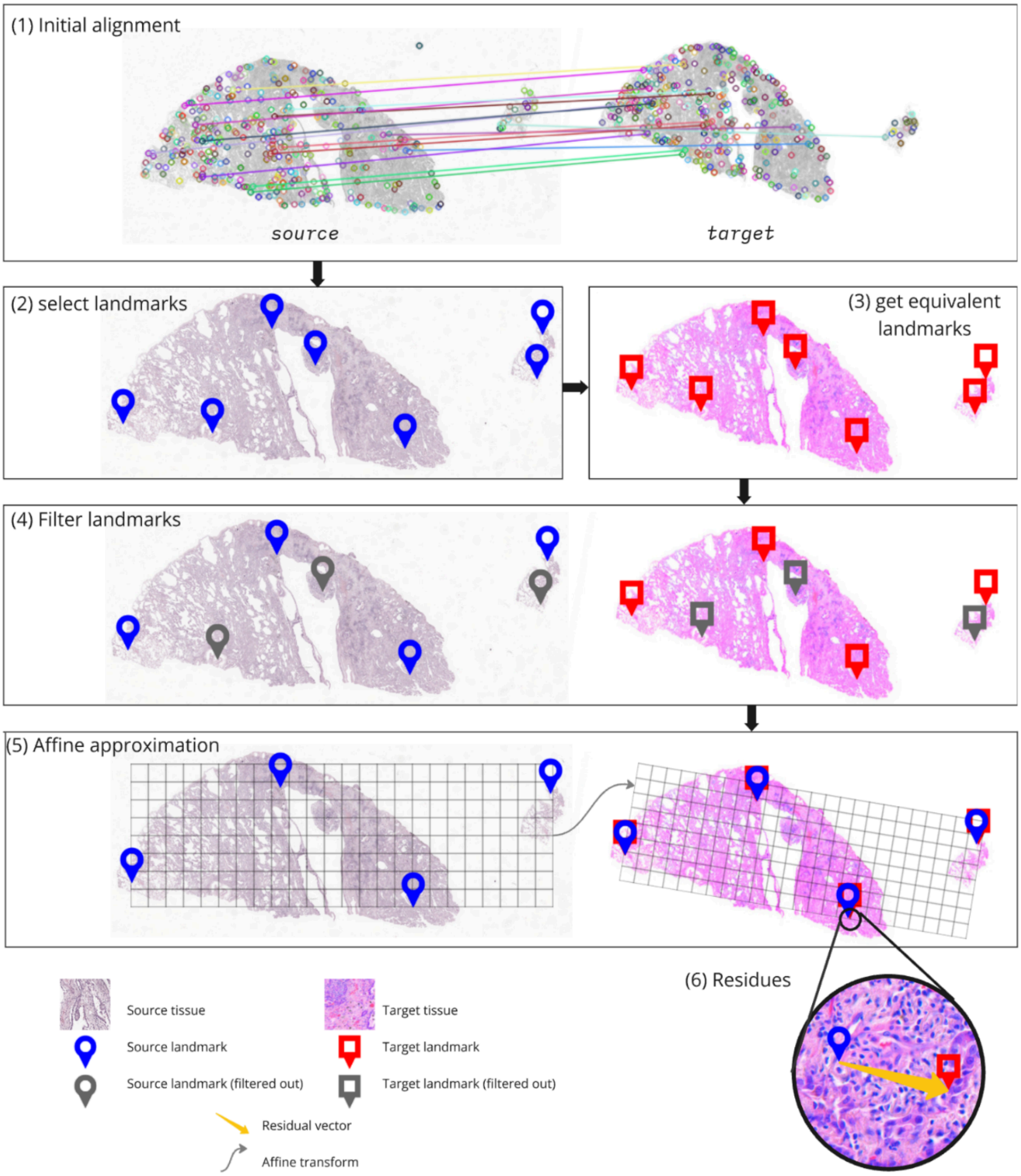

*Figure 2: Overview of the UWarp slide registration process.*

1. Find an initial transformation between the source and target slide at low resolution using SIFT-based feature matching.
2. Select N random landmarks on the source slide.
3. Perform iterative patch registration for each of these landmarks (more details on Figure 3).
4. Filter out the landmarks where the registration doesn't match perfectly.

5. Find the least square error affine transformation that aligns the source landmarks with the target landmarks.
6. Store residual vectors for local adjustments.

## 2.1 Initial transformation

In order to successfully warp coordinates from the source slide to the target slide, we first perform a low-resolution initial alignment of the source and target modalities. As done by [13], we determine an initial affine transformation by aligning key points extracted with the Scale Invariant Feature Transform (SIFT).

## 2.2 Landmarks definition

We automatically define uniformly distributed points called *landmarks* on the source slide. These points will later be used for Iterative Patch Registration (see section **3.Iterative Patch Registration** in Methods). It is crucial to note that only regions containing tissue can be successfully registered; regions consisting solely of background cannot undergo successful registration.

The following procedure outlines the methodology for defining a set of landmarks. It is also depicted on **Figure S1**:

1. Generate a thresholded version of the thumbnail image using Otsu [23] , where white pixels denote tissue.
2. Define a grid of squares of size S on the thumbnail surface.
3. For each square:
   - If the square contains tissue: randomly select one white pixel
   - If the square contains no tissue: continue to the next square

The choice of S is made automatically by the algorithm to produce approximately the number of landmarks requested by the user. This process involves two steps: First, S is initially set to an arbitrary value of 50, and the algorithm calculates how many landmarks this value produces. Then, S is adjusted to better match the requested number of landmarks To get the new value of S, we multiply 50 by $\sqrt{(produced_landmarks/requested_landmarks)}$ .

## 2.3 Iterative Patch Registration

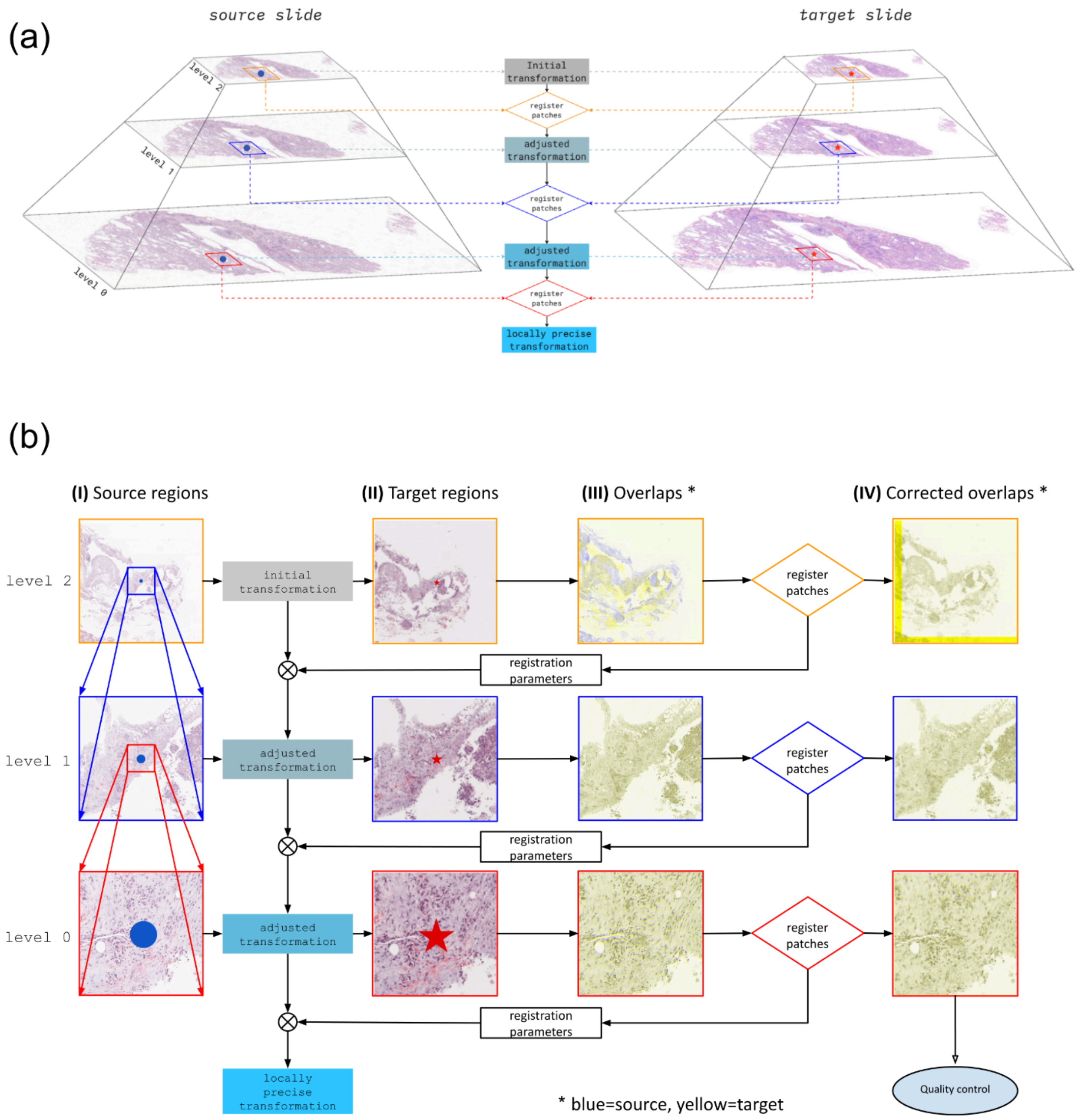

*Figure 3: Iterative Patch Registration (IPR) process.* ***(a)*** *General overview of IPR on different magnification levels.* ***(b)*** *Detailed view of tissue patches during IPR.*

In **Figure 3a**, we present an overview of Iterative Patch Registration (IPR), an iterative tissue alignment method that uses successive registration to gradually refine and adjust an affine transformation that warps points from the source slide to the target slide. We use the pyramidal nature of WSIs to get a precise affine transformation in the neighborhood of a given source point. IPR consists of the following steps, detailed further in the next paragraphs and in Figure 3b:

<u>Step 1: Define successive regions centered on the source slide landmark</u>

Iteratively define a sequence of regions on successive magnification levels such that they are all centered around the landmark, and have the same pixel dimensions. On **Figure 3b**, column **(I)** depicts the choice of 3 reference regions on each level of the source slide. Column **(II)** is the equivalent region on the target slide according to the current best known estimate of the transformation. **(III)** is the Blue-Yellow overlap of grayscale versions of **(I)** and **(II)**. After using the *elastix_registration_method* function of the ITK [24] library, we apply an adjustment to the coordinates of the target region so that the two regions overlap perfectly, as displayed in the adjusted overlap **(IV)**. This allows us to locally refine the approximation of our transformation for the next steps.

Step 2: Warp and adjust

Applying the previously calculated initial transformation (see section **1.Initial transformation** in Methods) to the coordinates of the first region, we get an equivalent region on the target slide. We perform a registration between the two regions using the *elastix_registration_method* function of the ITK library. The parameters of this registration step are used to enhance the local accuracy of the transformation function.

Step 3: Repeat until reaching level 0

We repeat step 2 for all of the regions defined in step 1, iteratively refining our approximation until we reach the highest resolution level. At the end of this procedure, we can retrieve the target region coordinates corresponding to the source region coordinates with sub-cellular scale accuracy.

## 2.4 Quality control: filtering out landmarks with poor registration

We assessed the success of IPR for a given landmark using the Normalized Mutual Information (NMI) metric, an established information-based similarity measure [25]. Registrations with NMI below an empirically determined threshold were considered flawed and thus discarded. This quality control mechanism is independent of ITK and ensures that only landmarks with high registration accuracy are retained in our workflow. Based on preliminary evaluation (see ***Results 4***), we set the NMI threshold at 0.15 to balance alignment quality and landmark retention.

## 2.5 Affine approximation using least squares

IPR is very precise but can take up to a few seconds per patch. It is not suited for systematic use in fast compute-time applications, such as deep learning pipelines. To address this, we rely on a single affine transformation for efficiently mapping equivalent patch coordinates between slides. Specifically, we determine the affine function that best describes the transformation between source and target landmarks, minimizing the least squares error using the least squares algorithm. Due to acquisition imperfections, local noise, and possible artifacts, precise alignment often requires multiple landmarks. Therefore, we use a larger set of points, evenly distributed across the tissue (see section **2. Landmarks definition** in Methods).

## 2.6 Non rigid transformation based on local adjustments

Due to the high complexity of biopsy preparation, some images may present artifacts like tissue rips or tissue folds. Furthermore, the image reconstruction algorithm that provides the WSI file may be subject to slight inaccuracies, producing tearing artifacts. Thus, it is often impossible to perform accurate region matching with a purely affine transformation. We need to add non-linearity (non-rigid transformations) to our transformation in order to account for these possible artifacts.

The least squares algorithm outputs a residual vector for each given landmark. It corresponds to the local error of the affine approximation. The better the alignment of the source and target landmarks, the smaller the norm of the residual vectors.

The residual vectors can be stored alongside the landmark coordinates and used to correct our affine transformation at runtime. Since there is only a limited number of landmarks on the slide, we define an interpolation method for the correction to be applied in between landmarks. We tested three error interpolation methods and compared them in Results **2. Selection of the best non-rigid adjustment method for UWarp**

- No correction: the residual vectors are ignored and only the affine transformation is used.
- Nearest neighbor: Apply the correction corresponding to the nearest residual.
- Linear interpolation: Apply a linear combination of the closest residuals.

## 2.7 Metric choice for performance evaluation

The scaling, shearing, and rotational components of the transformation are resolution-independent, making them straightforward to predict accurately. Consequently, errors at the patch level arise almost exclusively from the translational component of the transformation, as also noted by [13] .

To evaluate performance, we selected Target Registration Error (TRE) as our primary metric, as it directly measures the translational component of the transformation. TRE is computed using phase cross-correlation, a method that determines the translational pixel offset between two images. In this approach, the predicted and target regions are first transformed into the frequency domain using the Fourier transform. By analyzing the phase information alone, which encodes positional details, the relative shift between the two images is identified. The inverse Fourier transform of the normalized phase difference yields a sharp peak, whose position corresponds to the pixel displacement. This displacement provides a precise measurement of TRE. The p-values used to measure statistical significance were computed using a two-sided Mann–Whitney U test.

# 3.Results

The experimental process is structured into three key stages. First, we evaluate different parameter settings to identify the optimal balance between computational efficiency and accuracy for UWarp. Next, we compare the performance of UWarp against state-of-the-art registration methods to assess its effectiveness relative to existing approaches. Finally, we present an experimental characterization of scanner-induced local domain shift, demonstrating how our tool can be used to analyze and quantify these variations at a granular level.

## 3.1 Dataset description

We use two datasets composed of same-stain histological slides, scanned multiple times under varying conditions (e.g., different scanner models, brightness, and contrast settings). The dataset designed in the framework of the European project BosomShield, called the BosomShieldBreast dataset will be used to evaluate the impact of parameter changes in our method, as well as to demonstrate a concrete use case of domain shift analysis using the Breast-NEOprAIdict model. The CypathLung dataset will be used to compare performance against state-of-the-art methods, employing the same parameters optimized on the BosomShieldBreast dataset to ensure fair cross-dataset evaluation.

We use only same-stain scans because the objective is to characterize technical variability of histopathological samples, and reduce biological variability to a minimum. Additionally, the datasets feature slides with undetermined orientation, testing the algorithm's ability to align slides regardless of their initial positioning. Unlike annotated datasets such as ACROBAT [26] or ANHIR [27], these private datasets are not annotated, meaning we don't have a trivial way of computing TRE. This adds complexity to the development and evaluation of our registration methods (see section **7. Metric choice for performance evaluation** in Methods).

The CypathLung dataset is a private dataset composed of lung adenocarcinoma tissues, including 250 H&E physical slides that are scanned by 3 different scanners, providing 750 whole slide images. The BosomShieldBreast dataset is a breast cancer dataset composed of 50 H&E physical slides, scanned with 5 different scanners, providing 250 whole slide images.

| **Dataset** | CypathLung | | | BosomShieldBreast | | | | |
|---|---|---|---|---|---|---|---|---|
| **Usage** | Performance evaluation against state-of-the-art WSI registration techniques | | | Algorithm parameters tuning and real use case of domain shift study | | | | |
| **Number of samples** | 250 | | | 50 | | | | |
| **Scanner Name given** | C1 | C2 | C3 | B1 | B2 | B3 | B4 | B5 |
| **Vendor** | Leica | 3DHistech | Roche | Hamamatsu | Roche | 3DHistech | Leica | Hamamatsu |
| **Model** | Aperio GT450 | Mirax Scan | Ventana DP 200 | NanoZoomer-2.0 HT C9600 | Ventana DP200 | Mirax Scan | AT2 | NanoZoomer-2.0 HT C9600 |
| **Microns per pixel** | 0.264 | 0.121 | 0.250 | 0.227 | 0.465 | 0.243 | 0.500 | 0.454 |
| **File Format** | .svs | .mrxs | .bif | .ndpi | .tif | .mrxs | .svs | .ndpi |
| **Example Thumbnail** | 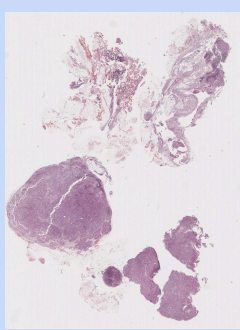 | 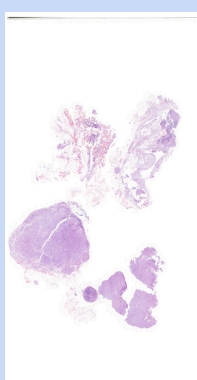 | 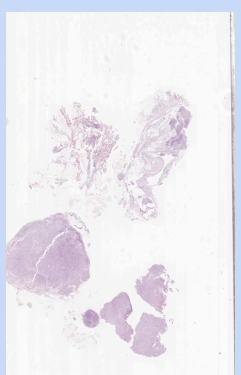 | 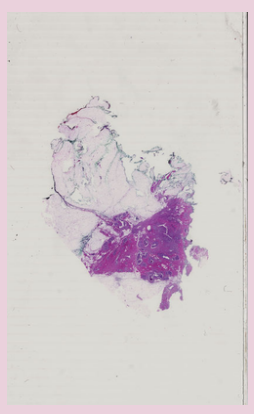 | 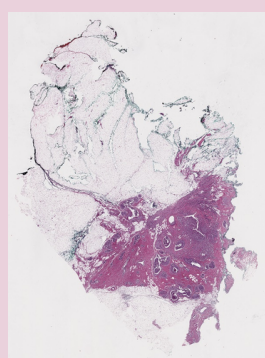 | 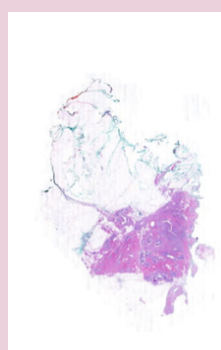 | 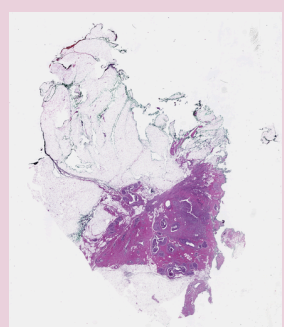 | 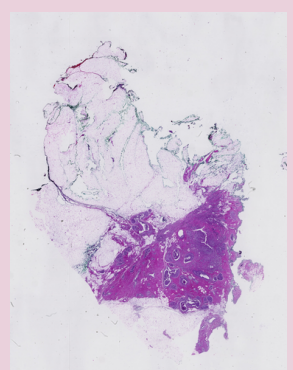 |

***Table 1**: Datasets, scanner properties and thumbnail examples*

## 3.2 Selection of the best non-rigid adjustment method for UWarp

The experiment was conducted on 50 slides of the BosomShieldBreast Dataset: 10 samples each scanned by 5 scanners (Refer to **Table 1** for scanner properties). We performed each registration with a landmark_number parameter of 30 following a cyclic registration scheme where slides were aligned in sequence from Scanner B1 to B2, B2 to B3, and so on, with B5 registering back to B1. We chose to evaluate median TRE error per slide as it is traditionally done in the state-of-the-art, but also include 95th percentile error to quantify the error in the worst case scenarios.

We denote on **Figure 4a** that the Nearest neighbor and Linear interpolation are superior to the No correction method ($p<0.001$ for both median and 95th percentile). The Linear interpolation performs better than the Nearest neighbor correction but only to a significant extent in our experiments for the median TRE ($p<0.001$ for median and $p=0.245$ for 95th percentile). We choose to keep the linear correction method for the rest of the experiments.

## 3.3 Selection of the optimal number of landmarks for UWarp

The main parameter in our algorithm is the number of landmarks that should be used for the affine approximation of the transformation. The time taken to compute a registration is proportional to the number of landmarks. In this experiment, identical registrations have been performed on the BosomShieldBreast dataset using 10, 30, or 100 landmarks. **Figure 4b** shows that the higher the number of landmarks, the lower the error, with the median of the median TRE per slide ranging from 2.0 pixels for 10 landmarks to 1.0 pixel for 100 landmarks, and the median of the 95th percentile TRE per slide ranging from 3.7 pixels for 10 landmarks, to 2.2 pixels for 100 landmarks. Using 100 landmarks provides an almost perfect alignment, with all the recorded registrations having a 95th percentile TRE lower than 4.0 pixels (about 1 micrometer at 40x magnification). Compute time being proportional to the number of landmarks, using 30 landmarks provides a good balance between compute time and accuracy and makes the registrations accurate enough for later use in downstream tasks such as deep learning pipelines.

## 3.4 Selection of the optimal NMI threshold for landmark filtration

To determine an appropriate threshold for the NMI metric in filtering registration quality, we conducted a systematic evaluation of patch-level alignments across a range of NMI values. Registration success was assessed visually based on overlap quality: a registration was considered successful when the source and target patches exhibited perfect alignment, resulting in a uniform grayscale appearance in the overlap image (as opposed to color artifacts indicating misalignment).

| NMI range | 0 < nmi < 0.10 | 0.10 < nmi < 0.20 | 0.20 < nmi |
|---|---|---|---|
| Alignment examples | 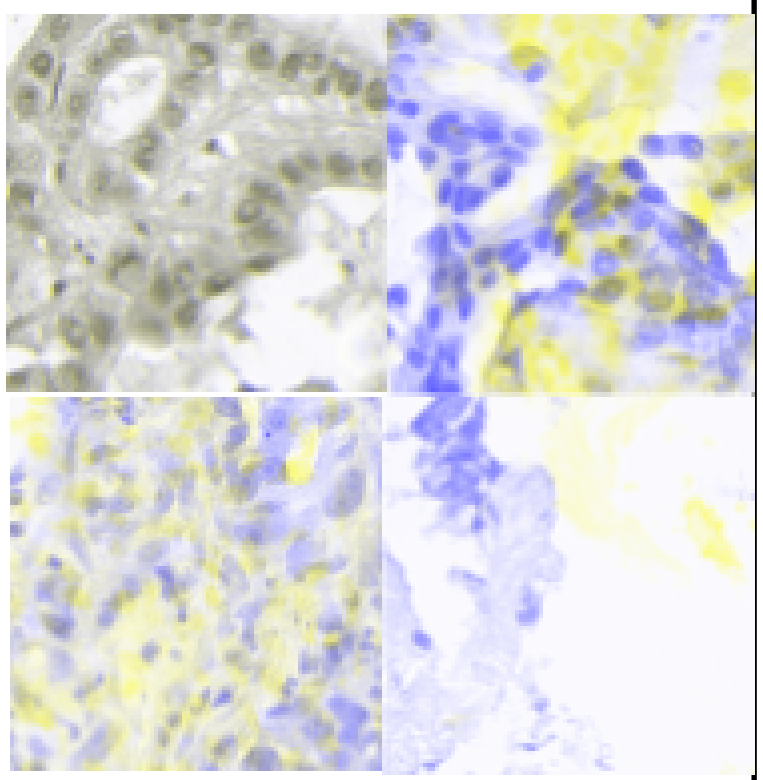 | 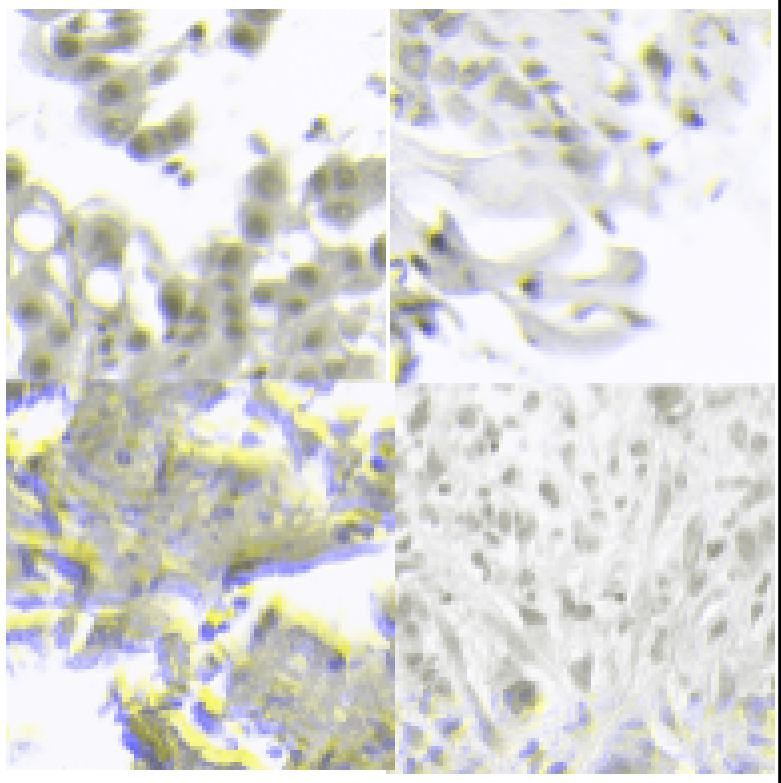 | 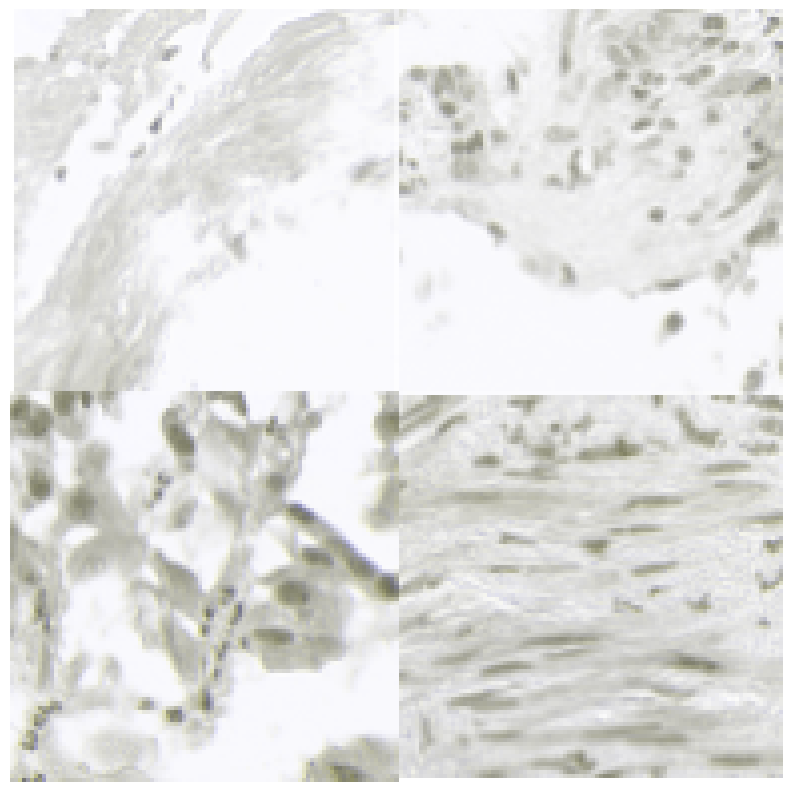 |
| Observed proportion of successful alignments | ~10 % | ~95% | >99% |

***Table 2***: *Effectiveness of the NMI metric for filtering out bad registrations.*

After careful consideration of these results, we chose to set the default NMI threshold at 0.15, providing a good balance between a high quality threshold and enough landmarks. During our experiments, this NMI value proved to be largely unaffected by the scanners used.

## 3.5 Performance of UWarp against the state-of-the-art

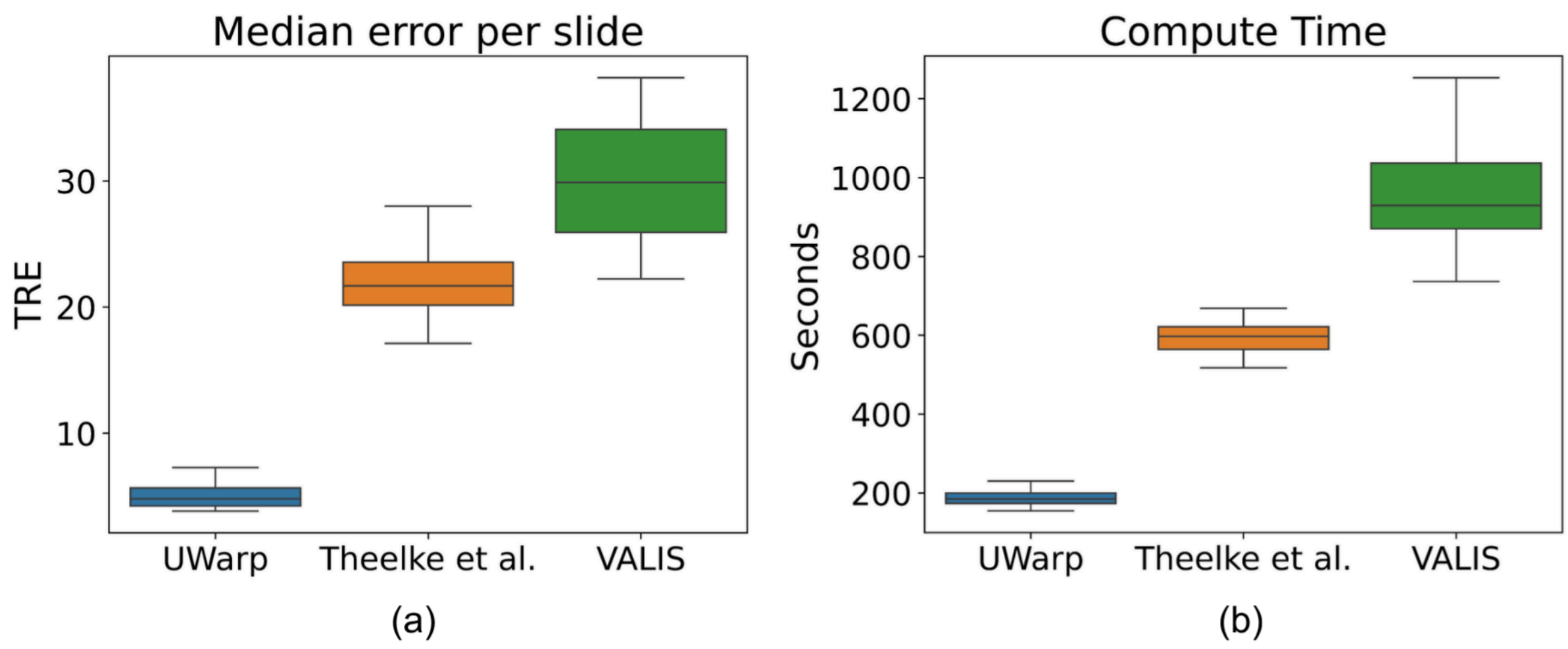

***Figure 4:*** *UWarp performance against the state-of-the-art.* ***(a)*** *Median Target Registration Error (in pixels) per registration* ***(b)*** *Compute time per registration.*

**Figure 4** compares UWarp with state-of-the-art registration algorithms on 500 registration tasks across the 250 slides of the CypathLung dataset, with half from the C1 to C2 scanner pair and half from the C1 to C3 pair. The implementation from Theelke et al. was run using the authors' default parameters. For VALIS, micro-registration was used with the highest resolution fraction (*0.1*) that could reliably run on our 64 GB RAM machine without crashing. As shown in **Figure 4a**, with 30 landmarks, UWarp achieved a median Target Registration Error (TRE) of 5 pixels (approximately 1.25 µm), compared to 22 pixels (approximately 5.5 µm) for Theelke et al. and 30 pixels (approximately 7.5 µm) for VALIS. All differences were statistically significant ($p < 0.001$). In terms of computational efficiency (**Figure 4b**), UWarp also outperformed both baselines significantly ($p < 0.001$), with a median runtime of 184 seconds per slide, compared to 597 seconds for Theelke et al. and 929 seconds for VALIS. **Figure S2** gives a visual comparison of the performance of our registration algorithm against the state-of-the-art, on patches of size 512x512 at the highest available magnification (0.25mpp).

## 3.6 Experimental Characterization of Scanner-Induced Local Domain Shift

The following section aims to instantiate the pipeline shown in **Figure 1** in order to characterize the scanner-induced domain shift of a binary classification model at a local scale. The binary classification model used for this study is Breast-NEOprAIdict [28], an advanced deep learning model designed to predict pathological complete response (pCR) in early breast cancer patients undergoing standard neoadjuvant chemotherapy. Leveraging data from the initial diagnostic biopsy, this model aims to enhance precision medicine by providing personalized insights into tumor chemosensitivity. By optimizing treatment strategies, Breast-NEOprAIdict holds the potential to significantly improve patient outcomes in breast cancer care. In this study, we do not wish to evaluate the reliability of Breast-NEOprAIdict, our only interest lies in the study of the prediction variability due to inter-scanner domain shift.

Following the process described in Figure 1, we use UWarp on each slide of the BosomShield Breast dataset (50 samples, 5 scans per sample). For each sample, we perform the following registrations:

- Scanner B1 to scanner B2
- Scanner B1 to scanner B3
- Scanner B1 to scanner B4
- Scanner B1 to scanner B5

Then for each sample, we take the B1 scan as a reference and we automatically select 100 random points of tissue (non-background) on it. On each point of tissue, we then create a source region of 512x512 pixels centered on this point. We warp this source region onto the B2, B3, B4 and B5 scans to retrieve the corresponding target regions. This process gives us 50x5x100=25000 patches of tissue. We then run Breast-NEOprAIdict on each of those patches For each patch, the model outputs a score between 0 and 1, corresponding respectively to low and high pCR response probability. We also generate slide-level predictions by using median aggregation, following the standard Breast-NEOprAIdict pipeline.

| Pearson R | B2 | B3 | B4 | B5 | Mean absolute difference | B2 | B3 | B4 | B5 |
|---|---|---|---|---|---|---|---|---|---|
| B1 | 0.88 (0.95) | 0.71 (0.80) | 0.87 (0.94) | 0.88 (0.93) | B1 | 0.05 (0.02) | 0.10 (0.09) | 0.05 (0.02) | 0.05 (0.03) |
| B2 | 1 | 0.63 (0.71) | 0.86 (0.95) | 0.85 (0.90) | B2 | 0 | 0.12 (0.11) | 0.06 (0.02) | 0.06 (0.05) |
| B3 | | 1 | 0.68 (0.73) | 0.74 (0.88) | B3 | | 0 | 0.11 (0.10) | 0.08 (0.06) |
| B4 | | | 1 | 0.82 (0.88) | B4 | | | 0 | 0.06 (0.04) |

***Table 3****: Pearson Correlation and Mean absolute prediction difference table for each pair of scanners at patch-level (n=5000) (at slide-level (n=50)).*

On **Table 3**, we computed the Pearson correlation coefficient and the mean absolute prediction difference for the set of patch-level and slide-level model predictions between each pair of scanners. We notice that some pairs of scanners show a higher level of agreement than others. 3DHistech shows lower overall correlation with other scanners. We also show that patch-level and slide-level correlation coefficients are strongly correlated, although the slide aggregation has an unforming impact, with Pearson's R coefficient being systematically higher at slide-level than at patch-level, and the mean absolute difference being systematically lower at slide-level than at patch-level.

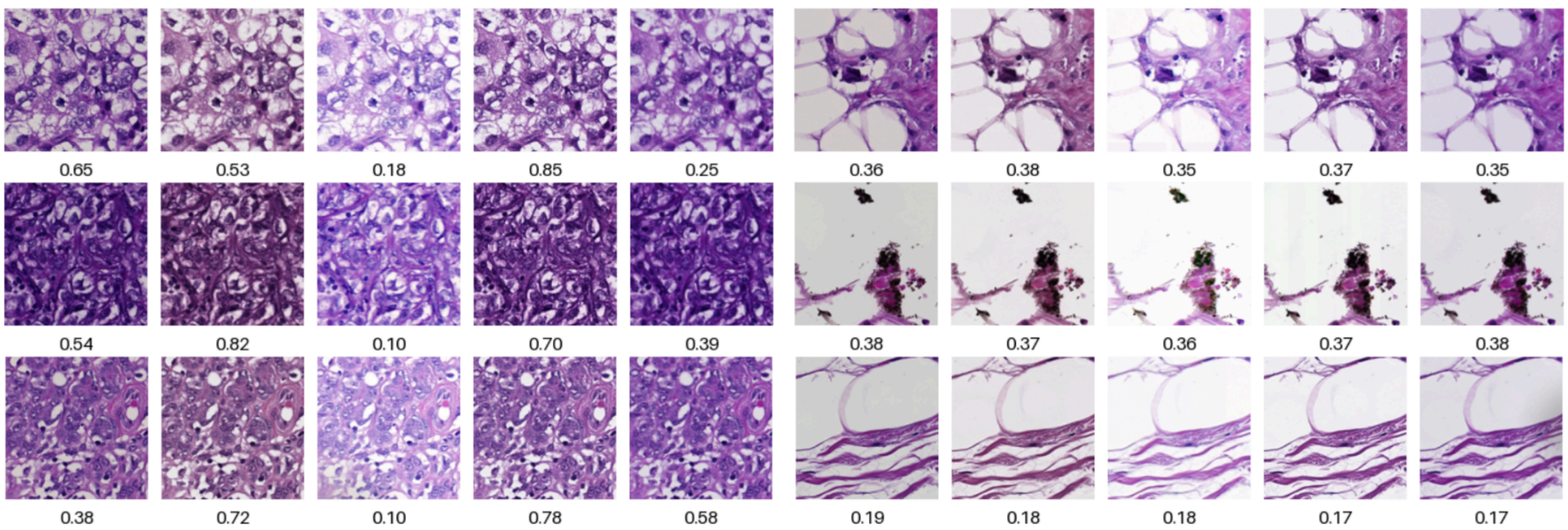

*Figure 5: Breast-NEOprAIdict [28] prediction scores across the 5 scanners (B1 to B5, left to right). The left-hand side corresponds to the highest discrepancies; the right-hand side corresponds to the lowest discrepancies.*

On **Figure 5,** we note that on this dataset, patches with high discrepancies tend to present tissue with higher cell density. Patches with lower discrepancies, on the contrary, tend to present less cell density, and proportionally more background. Importantly, while patches with high cell density also tend to exhibit higher overall prediction scores, this potential bias has been accounted for by focusing on relative variability rather than absolute values. Further examples are displayed **Figure S4** and **S5**.

These results demonstrate that the proposed registration-based pipeline enables precise, local-scale characterization of scanner-induced domain shift in model predictions. By isolating and comparing aligned tissue regions across scanner acquisitions, our method exposes variability that would otherwise be masked at the whole-slide level. Notably, we observe that certain scanners introduce more prediction drift than others, and that tissue characteristics (e.g., cell density) modulate the sensitivity to scanner effects. This underlines the importance of local domain shift analysis for validating model robustness and highlights the pipeline's potential as a practical tool for quality assurance, scanner benchmarking, and pre-deployment evaluation in digital pathology workflows.

# 4.Discussion

UWarp is introduced as a novel slide registration pipeline specifically designed to tackle scanner-induced domain shift in computational histopathology. UWarp combines coarse global alignment with fine-grained local adjustments to achieve accurate and robust registration of tissue patches between different scans. This tool builds the foundation for further investigation on local domain shift. Unlike traditional registration methods that aim for seamless whole-image reconstruction, UWarp performs patch coordinate warping to enable a finer, localized analysis of prediction variability.

UWarp combines a coarse global alignment (using fast linear approximations) with fine-grained, multi-level pyramidal adjustments to generate an accurate deformation field. We iteratively apply affine patch-level corrections to create landmarks for local adjustments, whereas the state-of-the-art only uses translational patch-level adjustments. The pipeline is robust to moderate misalignments at the initial stage because subsequent iterative transformations progressively refine the registration. Furthermore, the design is flexible: while the paper illustrates square patches for simplicity, the approach applies to any rectangular source patch shape. Although denser landmark placement increases computational cost, it provides a more accurate local interpolation. The reliability of IPR-produced landmark pairs is monitored using a quality control metric, normalized mutual information (NMI), which is chosen for its independence from spatial features, unlike internal ITK metrics. An NMI threshold around 0.15 was found to be optimal across multiple scanners, although the metric must be balanced to avoid either passing unsuccessful registrations or compromising the density of local residual corrections. UWarp achieves a median TRE of less than 4 pixels (translating to <1 µm at typical 40× magnification) with a median execution time of 240 seconds, making it practical for large-scale domain shift analysis.

In evaluations involving 40 whole-slide images from five different scanners, strong correlations were observed between scanner pairs (patch-level correlations ranged from 0.63 to 0.88 and slide-level from 0.71 to 0.95). These results indicate that even validated algorithms can be significantly affected by inter-scanner variability. The data-driven insights provided by UWarp underscore the importance of examining domain shift at a local level, as well as the need for robust aggregation methods to improve model generalizability. The work further hypothesizes that domain shift is more pronounced in patches with higher cell density, lower amounts of background, and that the prediction differences from one scanner to another may be modeled by an affine transformation. Retrieving these affine coefficients might offer a calibration strategy to adjust predictions for a new target domain. Moreover, the framework opens the possibility of generating naturally augmented patches for data augmentation, providing high-quality natural positive pairs for self-supervised contrastive learning, without relying on synthetic data augmentation.

Scanner software malfunctions can lead to tearing artifacts like misstitching that produces non-tissue features as depicted on **Figure S6**. Such artifacts not only bias model predictions but also complicate affine-based registration by introducing non-rigid deformations. UWarp mitigates these issues by incorporating local adjustments that compensate for tearing, as demonstrated by comparing registration results with different landmark densities. Increased landmarks improve the interpolation of residual vectors, thereby capturing the true reconstruction shift more precisely.

The current study is limited to binary classification for breast cancer tissue and focuses on prediction variability rather than final classification performance. A comprehensive assessment would require an evaluation of how this local shift impacts end-to-end clinical performance.

# 5.Conclusion

We present a patch-alignment approach for studying patch-level domain shift using our novel registration framework named UWarp, which outperforms the state-of-the-art by a large margin. Based on our experiments, it can be concluded that the study and characterization of localized domain shift can be more deeply understood with our method. Our investigation has shown that different tissue types and background proportions are subject to different degrees of domain shift for a trained binary classification model. The proposed methodology based on UWarp can be readily used in practice, and is applicable to other datasets, models, and could potentially be used for different tasks, such as natural data augmentation, or creation of positive pairs for self supervised learning.

Further investigation is needed to deeply characterize domain shift at a local scale and to determine whether specific regions of a slide are inherently more susceptible to scanner-induced variations. Future research should not only extend this framework to additional tissue types but also explore diverse prediction tasks and validate the calibration of affine coefficients across a broader range of scanners.

Looking ahead, UWarp's precise local registration framework offers promising applications beyond domain shift analysis. By quantifying scanner-induced variability at a fine-grained level, it can serve as a powerful tool for scanner benchmarking and clinical quality assurance (QA), flagging devices or settings that introduce excessive distortion. Additionally, the extraction of affine transformation parameters between domains opens the door to systematic model calibration, enabling algorithms to adjust their predictions dynamically when deployed on new scanners. Moreover, UWarp's ability to generate pairs of naturally augmented tissue patches provides high-quality inputs for self-supervised contrastive pretraining, offering a scalable alternative to synthetic data augmentation and potentially boosting model robustness across diverse clinical environments.

# Code Availability

The UWarp registration pipeline developed for this study is available as open-source software on GitHub: *https://github.com/antoineschieb/UWarp*. The repository contains the full implementation and installation instructions. Please note that the datasets used to produce the results in this paper are not publicly available. The code is released under the MIT license, and contributions from the community are welcome.

# Funding

This work was supported by BosomShield, a project that has received funding from the Marie Skłodowska-Curie Doctoral Networks Actions (HORIZON-MSCA-2021-DN-01-01) under grant agreement 101073222. Further support was provided by SCARLET, a project funded through the Proyectos Estratégicos Orientados a la Transición Ecológica y a la Transición Digital, 2021 call from the Ministerio de Ciencia e Innovación, under grant number TED2021-130081B-C22, with funding from NextGenerationEU.

# Bibliography

1. Tellez D, Balkenhol M, Otte-Holler I, et al. Whole-Slide Mitosis Detection in H&E Breast Histology Using PHH3 as a Reference to Train Distilled Stain-Invariant Convolutional Networks. *IEEE Trans Med Imaging*. 2018;37(9):2126-2136. doi:10.1109/TMI.2018.2820199

2. Tschuchnig ME, Oostingh GJ, Gadermayr M. Generative Adversarial Networks in Digital Pathology: A Survey on Trends and Future Potential. Published online May 7, 2020. doi:10.48550/arXiv.2004.14936

3. Anghel A, Stanisavljevic M, Andani S, et al. A High-Performance System for Robust Stain Normalization of Whole-Slide Images in Histopathology. *Front Med*. 2019;6:193. doi:10.3389/fmed.2019.00193

4. Hoque MdZ, Keskinarkaus A, Nyberg P, Mattila T, Seppänen T. Whole slide image registration via multi-stained feature matching. *Comput Biol Med*. 2022;144:105301. doi:10.1016/j.compbiomed.2022.105301

5. Michielli N, Caputo A, Scotto M, et al. Stain normalization in digital pathology: Clinical multi-center evaluation of image quality. *J Pathol Inform*. 2022;13:100145. doi:10.1016/j.jpi.2022.100145

6. Runz M, Rusche D, Schmidt S, Weihrauch MR, Hesser J, Weis CA. Normalization of HE-stained histological images using cycle consistent generative adversarial networks. *Diagn Pathol*. 2021;16(1):71. doi:10.1186/s13000-021-01126-y

7. Nisar Z, Vasiljević J, Gançarski P, Lampert T. Towards Measuring Domain Shift in

Histopathological Stain Translation in an Unsupervised Manner. In: *2022 IEEE 19th International Symposium on Biomedical Imaging (ISBI)*. ; 2022:1-5. doi:10.1109/ISBI52829.2022.9761411

8. Stacke K, Eilertsen G, Unger J, Lundstrom C. Measuring Domain Shift for Deep Learning in Histopathology. *IEEE J Biomed Health Inform*. 2021;25(2):325-336. doi:10.1109/JBHI.2020.3032060

9. Stacke K, Eilertsen G, Unger J, Lundström C. A Closer Look at Domain Shift for Deep Learning in Histopathology. Published online September 26, 2019. doi:10.48550/arXiv.1909.11575

10. Ochi M, Komura D, Onoyama T, et al. Registered multi-device/staining histology image dataset for domain-agnostic machine learning models. *Sci Data*. 2024;11(1):330. doi:10.1038/s41597-024-03122-5

11. Kajihara T, Funatomi T, Makishima H, et al. Non-rigid registration of serial section images by blending transforms for 3D reconstruction. *Pattern Recognit*. 2019;96:106956. doi:10.1016/j.patcog.2019.07.001

12. Kiemen AL, Braxton AM, Grahn MP, et al. CODA: quantitative 3D reconstruction of large tissues at cellular resolution. *Nat Methods*. 2022;19(11):1490-1499. doi:10.1038/s41592-022-01650-9

13. Theelke L, Wilm F, Marzahl C, et al. Iterative Cross-Scanner Registration for Whole Slide Images. In: *2021 IEEE/CVF International Conference on Computer Vision Workshops (ICCVW)*. IEEE; 2021:582-590. doi:10.1109/ICCVW54120.2021.00071

14. Lotz J, Weiss N, Heldmann S. Robust, fast and accurate: a 3-step method for automatic histological image registration. Published online March 29, 2019. doi:10.48550/arXiv.1903.12063

15. Lotz J, Olesch J, Muller B, et al. Patch-Based Nonlinear Image Registration for Gigapixel Whole Slide Images. *IEEE Trans Biomed Eng*. 2016;63(9):1812-1819. doi:10.1109/TBME.2015.2503122

16. Wodzinski M, Müller H. DeepHistReg: Unsupervised Deep Learning Registration Framework for Differently Stained Histology Samples. *Comput Methods Programs Biomed*. 2021;198:105799. doi:10.1016/j.cmpb.2020.105799

17. Lowe DG. Distinctive Image Features from Scale-Invariant Keypoints. *Int J Comput Vis*. 2004;60(2):91-110. doi:10.1023/B:VISI.0000029664.99615.94

18. Hu Y, Modat M, Gibson E, et al. LABEL-DRIVEN WEAKLY-SUPERVISED LEARNING FOR MULTIMODAL DEFORMABLE IMAGE REGISTRATION. Published online 2018.

19. Wodzinski M, Marini N, Atzori M, Müller H. RegWSI: Whole slide image registration using combined deep feature- and intensity-based methods: Winner of the ACROBAT 2023 challenge. *Comput Methods Programs Biomed*. 2024;250:108187. doi:10.1016/j.cmpb.2024.108187

20. Jiang J, Larson NB, Prodduturi N, Flotte TJ, Hart SN. Robust hierarchical density estimation and regression for re-stained histological whole slide image co-registration. Sarder P, ed. *PLOS ONE*. 2019;14(7):e0220074. doi:10.1371/journal.pone.0220074

21. Rossetti BJ, Wang F, Zhang P, Teodoro G, Brat DJ, Kong J. Dynamic registration for gigapixel serial whole slide images. In: *2017 IEEE 14th International Symposium on Biomedical Imaging (ISBI 2017)*. IEEE; 2017:424-428. doi:10.1109/ISBI.2017.7950552

22. Gatenbee CD, Baker AM, Prabhakaran S, et al. Virtual alignment of pathology image series for multi-gigapixel whole slide images. *Nat Commun*. 2023;14(1):4502. doi:10.1038/s41467-023-40218-9

23. Otsu. A Threshold Selection Method from Gray-Level Histograms | IEEE Journals & Magazine | IEEE Xplore. Accessed March 24, 2025. https://ieeexplore.ieee.org/document/4310076

24. ITK: enabling reproducible research and open science - PubMed. Accessed March 26, 2025. https://pubmed.ncbi.nlm.nih.gov/24600387/

25. Shannon CE. A Mathematical Theory of Communication.

26. Weitz P, Valkonen M, Solorzano L, et al. ACROBAT -- a multi-stain breast cancer histological whole-slide-image data set from routine diagnostics for computational pathology. Published online November 24, 2022. doi:10.48550/arXiv.2211.13621

27. Borovec J, Kybic J, Arganda-Carreras I, et al. ANHIR: Automatic Non-Rigid Histological Image Registration Challenge. *IEEE Trans Med Imaging*. 2020;39(10):3042-3052. doi:10.1109/TMI.2020.2986331

28. Valderrama NF, Morel LO, Mweze DT, et al. Breast-NEOprAIdict: a deep learning

solution for predicting pathological complete response on biopsies of breast cancer patients treated with neoadjuvant chemotherapy. Published online December 17, 2024. doi:10.21203/rs.3.rs-5327033/v1

# Supplementary material

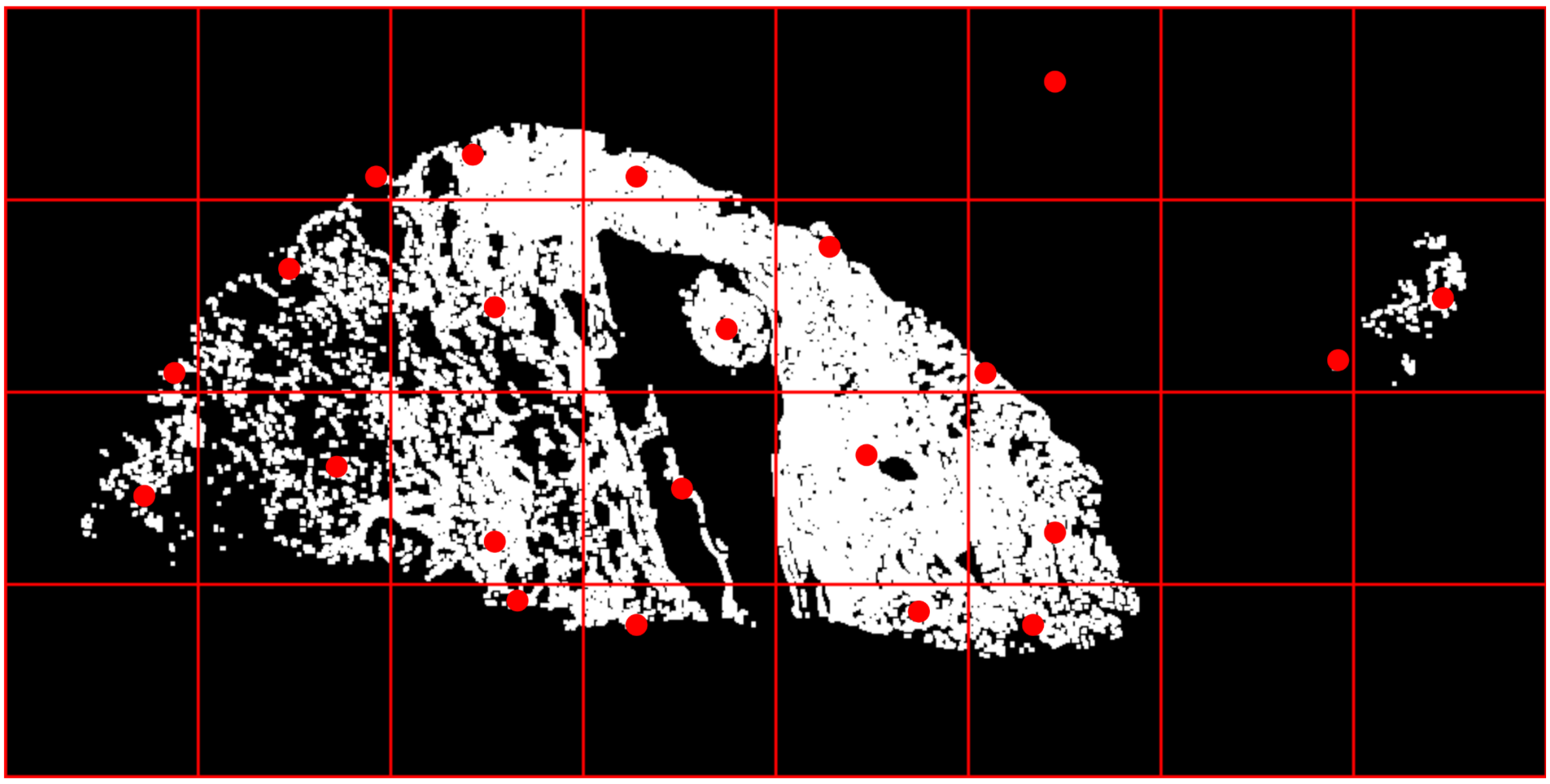

***Figure S1****: Example where S=100 on a 800x400 thumbnail. This choice of S yields 22 evenly distributed landmarks for this sample.*

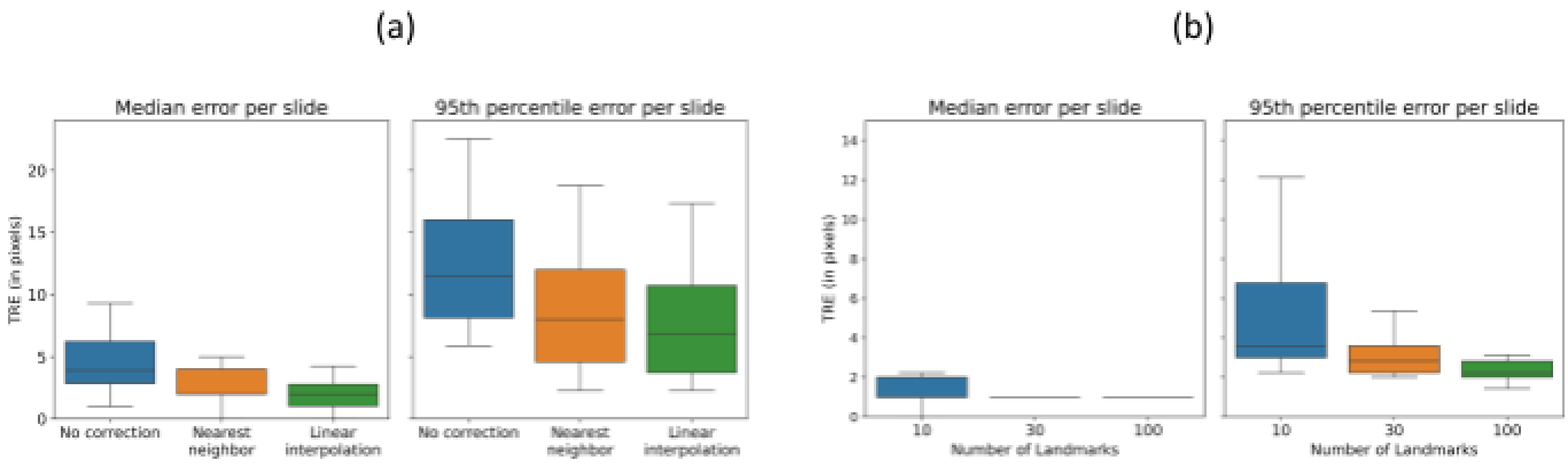

***Figure S2****: Results of the experimental process.* ***(a)*** *Error for each type of correction applied (n=50).* ***(b)*** *Error for each number of landmarks used (n=25).*

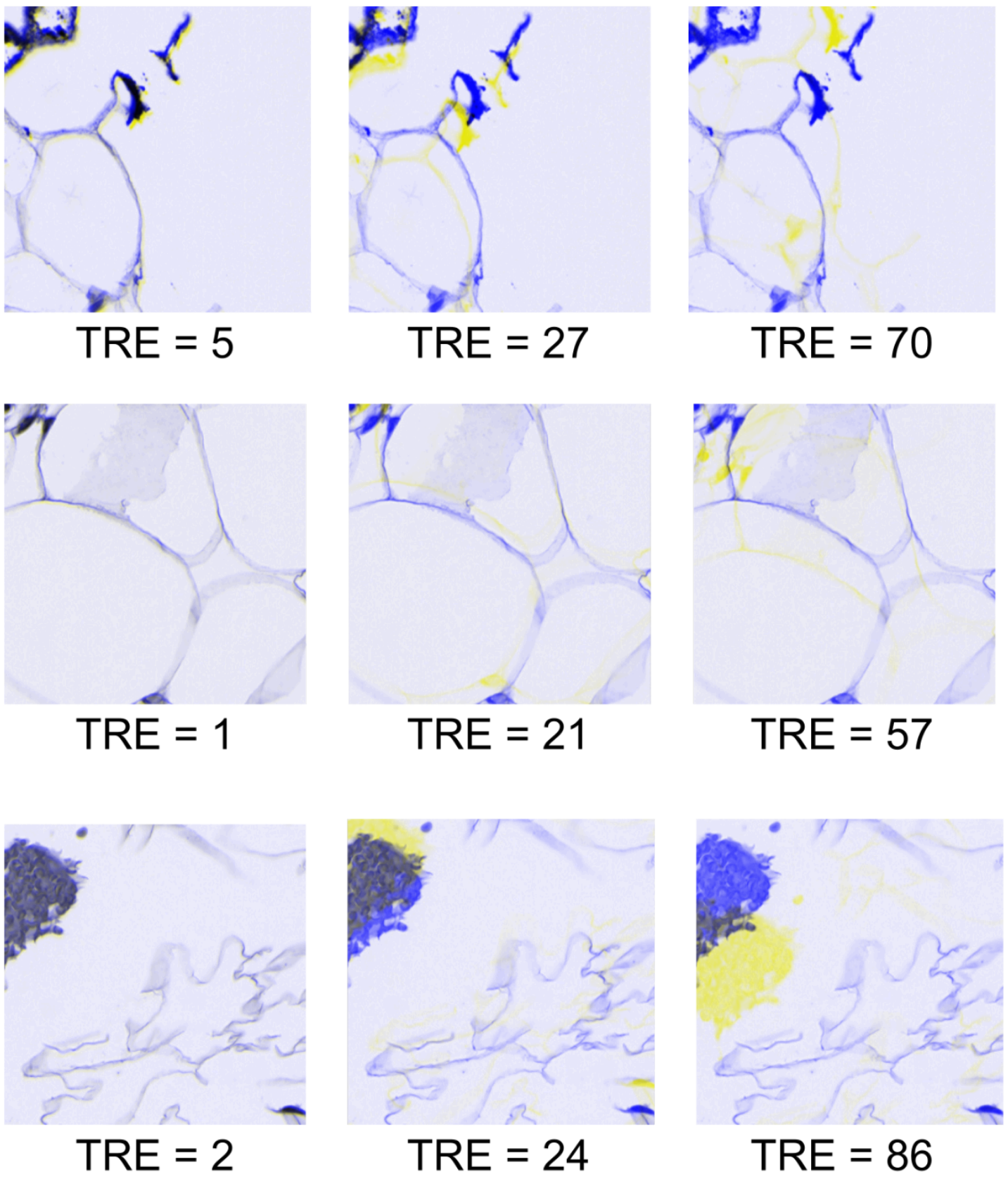

***Figure S3****: TRE (in pixels) and Blue-Yellow overlap of grayscale versions of source and target patches using transformations from three algorithms (Left: Ours, Middle: Theelke et al., Right: VALIS). If the tissue appears gray, it means that the patches are perfectly aligned.*

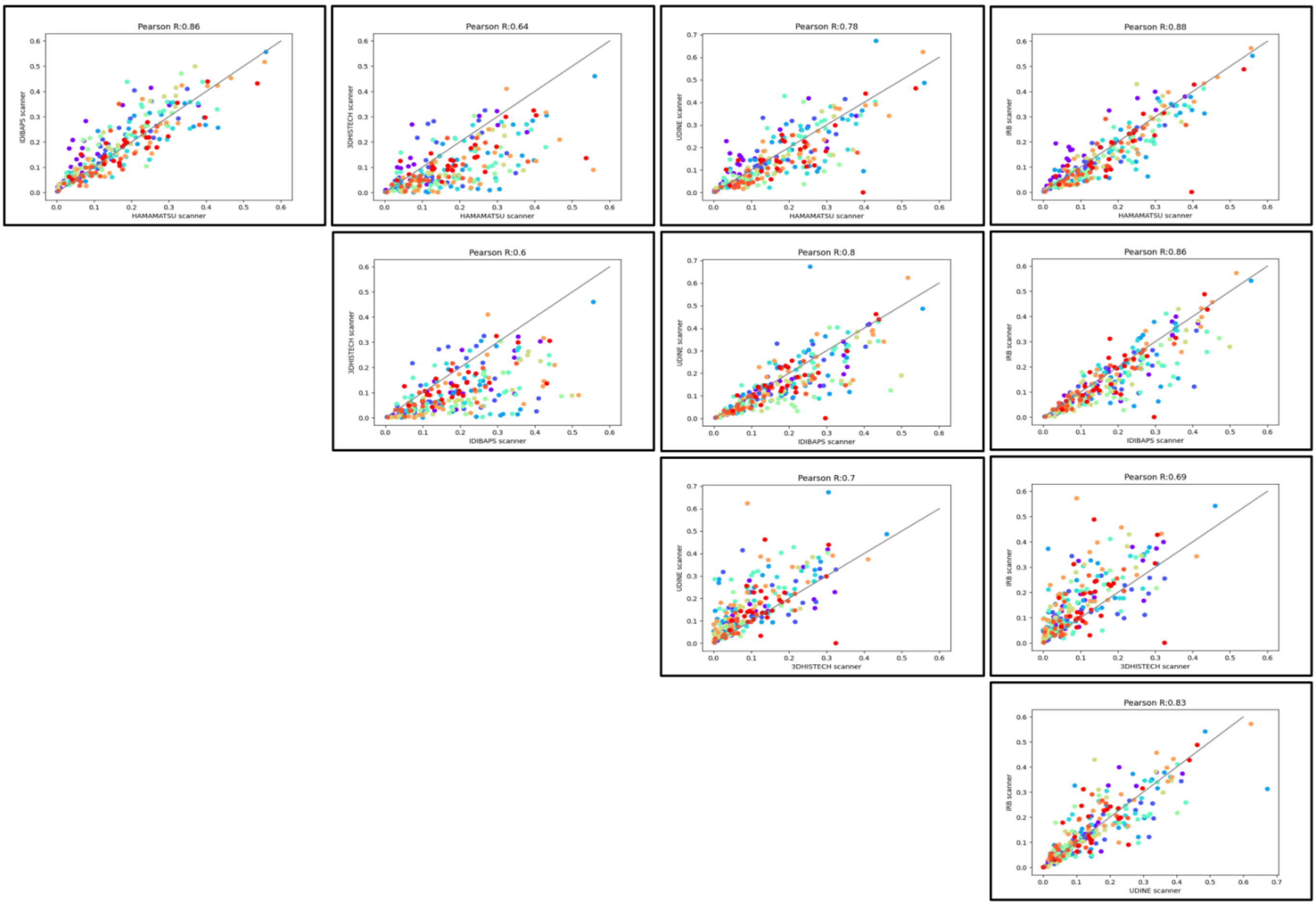

***Figure S4**: Correlation plots between each pair of scanners. Each point corresponds to a patch, and each color corresponds to one slide.*

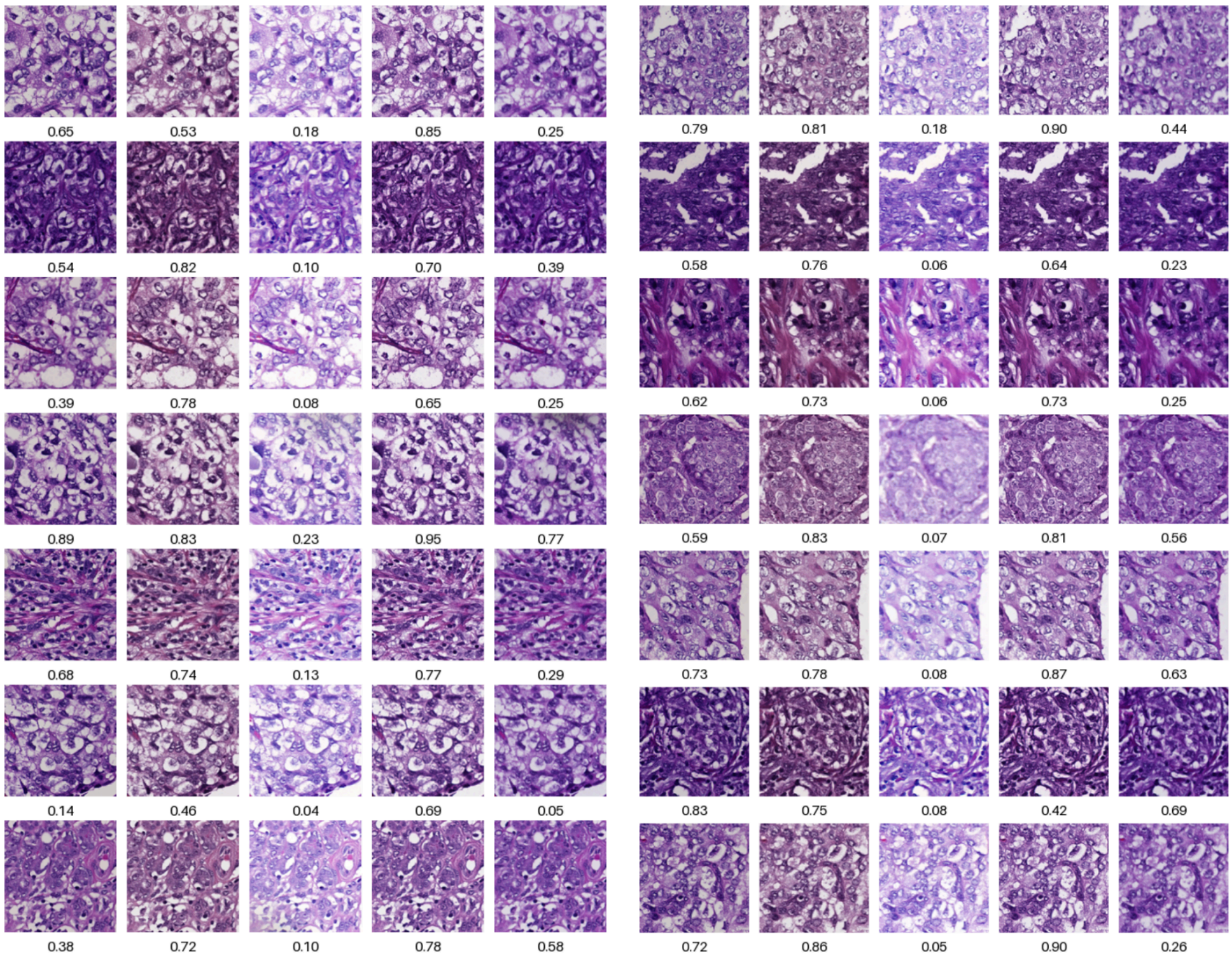

***Figure S5**: Patches with highest discrepancies in predictions  across scanners*

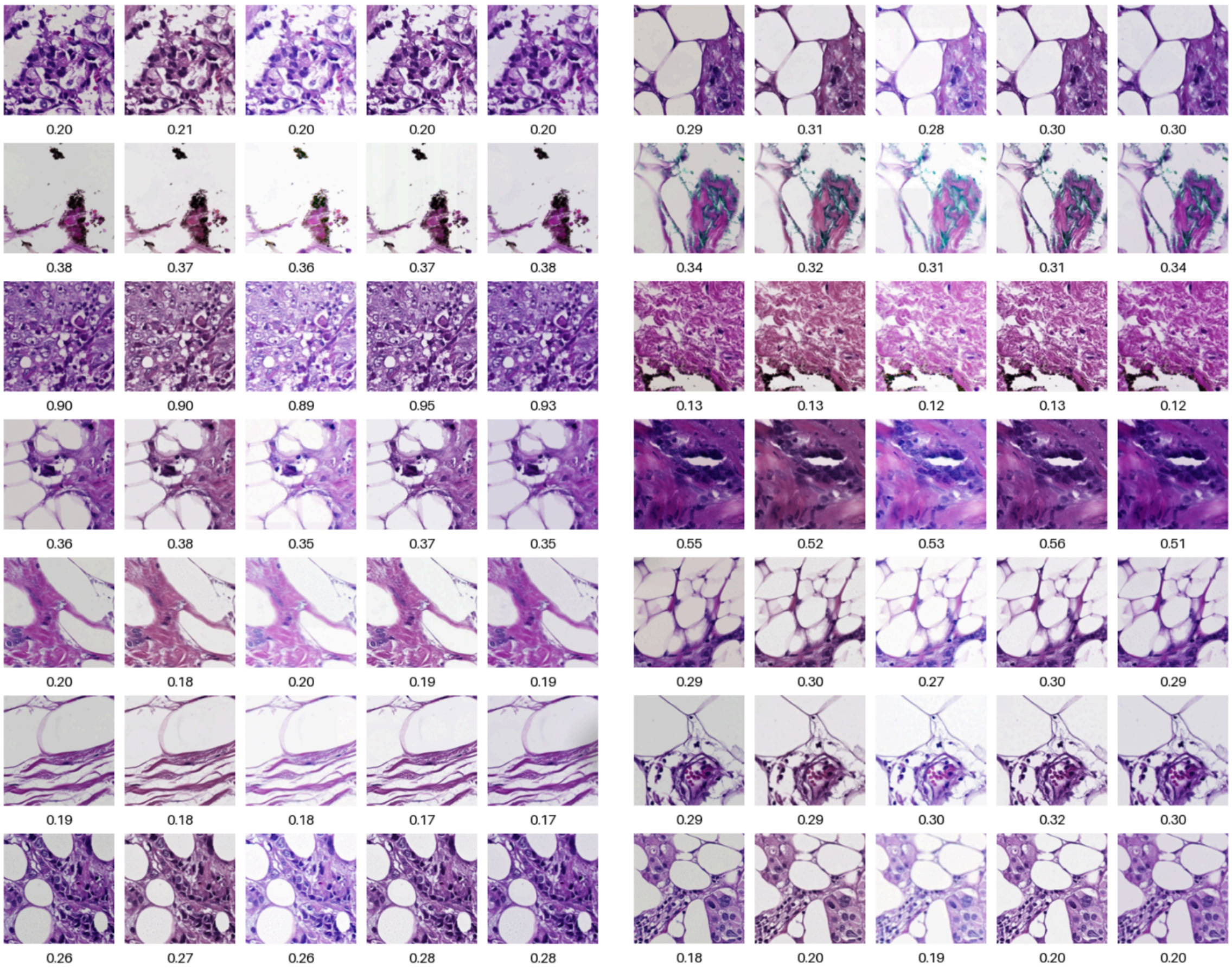

***Figure S6***: *Patches with lowest discrepancies in predictions across scanners*

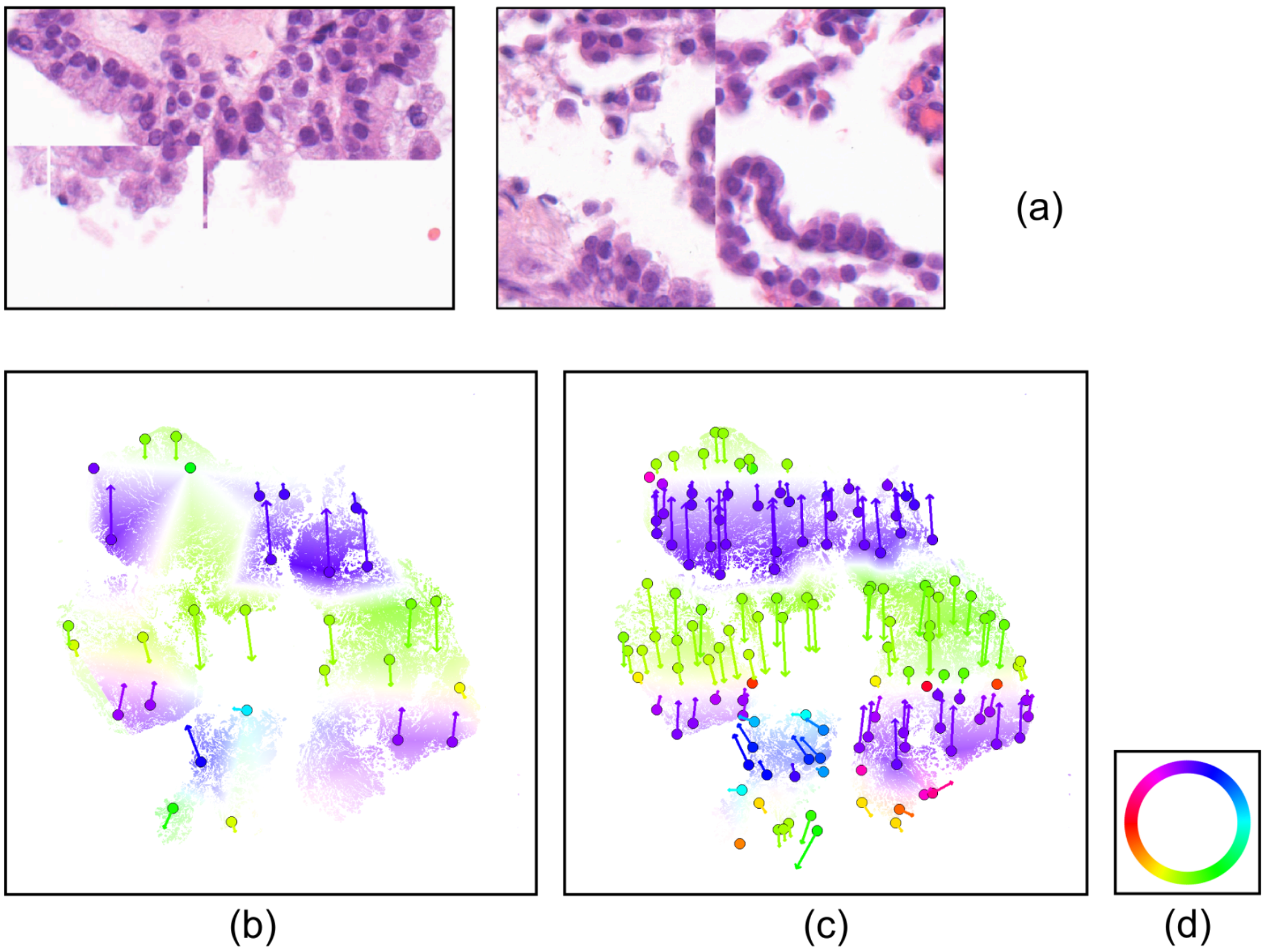

***Figure S7:*** *Vector-color representation of the residual vectors across the slide presenting reconstruction artifacts (residual vectors not to scale). The angle of the local adjustment vectors correspond to the hues presented on the legend* ***(c)****. The norm of the interpolation vector corresponds to the color saturation of the tissue.*